\documentclass[aip,pop,graphicx,amsmath, reprint]{revtex4-1}

\usepackage{siunitx}
\DeclareSIUnit\gauss{G}
\usepackage{lineno}
\usepackage{graphicx}
\usepackage{comment}
\usepackage{amssymb}
\usepackage[normalem]{ulem}
\begin{document}


\title{Ring momentum distributions as a general feature of Vlasov dynamics in the synchrotron dominated regime} 



\author{P. J. Bilbao}
 \email{pablojbilbao@tecnico.ulisboa.pt}
\affiliation{ 
GoLP/Instituto de Plasmas e Fus\~{a}o Nuclear, Instituto Superior T\'{e}cnico, Universidade de Lisboa, 1049-001 Lisbon, Portugal
}%

\author{R. J. Ewart}
\affiliation{%
Rudolf Peierls Centre for Theoretical Physics, University of Oxford, Oxford, OX1 3PU, UK
}
\affiliation{Balliol College, Oxford, OX1 3BJ, UK}
\author{F. Assunçao}
\affiliation{ 
GoLP/Instituto de Plasmas e Fus\~{a}o Nuclear, Instituto Superior T\'{e}cnico, Universidade de Lisboa, 1049-001 Lisbon, Portugal
}%
\author{T. Silva}%
\affiliation{ 
GoLP/Instituto de Plasmas e Fus\~{a}o Nuclear, Instituto Superior T\'{e}cnico, Universidade de Lisboa, 1049-001 Lisbon, Portugal
}%

 \author{L. O. Silva}%
  \email{luis.silva@tecnico.ulisboa.pt}
  \affiliation{ 
GoLP/Instituto de Plasmas e Fus\~{a}o Nuclear, Instituto Superior T\'{e}cnico, Universidade de Lisboa, 1049-001 Lisbon, Portugal
}

\date{\today}

\begin{abstract}
We study how radiation reaction leads plasmas initially in kinetic equilibrium to develop features in momentum space, such as anisotropies and population inversion, resulting in a ring-shaped momentum distribution that can drive kinetic instabilities. We employ the Landau--Lifshiftz radiation reaction model for a plasma in a strong magnetic field, and we obtain the necessary condition for the development of population inversion, we show that isotropic Maxwellian and Maxwell--J\"uttner plasmas, with thermal temperature $T>m_e c^2/\sqrt{3}$, will develop a ring-like momentum distribution. The timescales and features for forming ring-shaped momentum distributions, the effect of collisions and non-uniform magnetic fields are disscussed, and compared with typical astrophysical and laboratory plasmas parameters. Our results show the pervasiveness of ring-like momentum distribution functions in synchrotron dominated plasma conditions.
\end{abstract}

\pacs{}

\maketitle 

\section{Introduction}
The interplay between quantum electrodynamics (QED) and collective plasma dynamics has recently garnered significant interest \cite{grismayer2016laser,liseykina2016inverse, gong2019radiation, blackburn2020radiation, zhang2020relativistic, gonoskov2021charged, qu2021collective,qu2021signature, comisso2021pitch, bilbao2022radiation, zhdankin2022synchrotron, griffith2023radiation, qu2023creating, bulanov2024energy}. This is motivated by the prospects of directly achieving such regimes in the laboratory with the advent of higher-intensity lasers \cite{di2009strong, di2012extremely, thomas2012strong, vranic2014all, vranic2016quantum_m, fedotov2023advances, griffith2023radiation, bulanov2024energy}, magnetic-field amplification setups \cite{nakamura2018record, murakami2020generation, jiang2021magnetic}, and fusion plasmas \cite{hirvijoki2015radiation, decker2016numerical}. Conversely, astrophysical plasmas, such as those ubiquitous around compact objects, already have the necessary conditions for the interplay of collective plasma dynamics with QED effects \cite{turolla2015magnetars, kaspi2017magnetars, cerutti2017electrodynamics, xue2019magnetar,uzdensky2019extreme}. Some of the most fundamental QED processes involve photon emission. In strong electromagnetic fields (laser intensities of $10^{23}\, \SI{}{\watt/\centi\metre^2}$\ \ \cite{yoon2021realization} and magnetic-field strengths of $10^9\,\SI{}{\gauss}$\ \ \cite{jiang2021magnetic}), relativistic charged particles can radiate photons with energy comparable to the kinetic energy of the particle $(\gamma-1) m_e c^2$, where $\gamma$ is the Lorentz factor of the charged particle, $m_e$ is the electron mass, and $c$ is the speed of light in vacuum. In these scenarios, radiation reaction (i.e., the momentum recoil due to the radiation emission \cite{landau1975classical}) must be considered, modifying the dynamics of relativistic charged particles \cite{blackburn2020radiation, gonoskov2021charged}, and a significant fraction of the kinetic energy of the plasma can be transferred to synchrotron radiation.

Recent theoretical and numerical results have shown that radiation reaction induces ``bumps'' along the runaway-electron tail of these energetic particles in collisional fusion plasmas \cite{hirvijoki2015radiation, decker2016numerical}, produces phase-space attractors \cite{gong2016radiation}, enhances anisotropic acceleration in radiatively cooled plasma turbulence \cite{comisso2021pitch}, and efficiently drives  kinetic instabilities such as the firehose \cite{zhdankin2022synchrotron} and the electron cyclotron maser \cite{bilbao2022radiation}. These results hint at the importance of developing a deeper understanding of radiatively cooled plasma from a first-principles kinetic description.

Radiation reaction is not a conservative force; thus, it does not conserve the phase-space volume; moreover, it naturally generates anisotropies in momentum space \cite{bilbao2022radiation,zhdankin2022synchrotron}, as the cooling rates may differ along different directions depending on the electromagnetic-field configuration. This is evident for the case of synchrotron cooling in a constant magnetic field. As individual particles gyrate within the strong field, they emit synchrotron radiation, losing kinetic energy in the process. For ultra-strong magnetic fields, the radiated energy can be comparable to the initial kinetic energy of the particles. Consequently, an electron plasma will lose thermal energy through collective synchrotron emission, leading to a gradual cooling effect. 

The synchrotron radiative power is $P \propto \gamma^4 (\mathbf{p} \times \mathbf{a})^2$, where $\gamma$, $\mathbf{p}$ and $\mathbf{a}$ are the particle's Lorentz factor, momentum and acceleration, respectively \cite{lightman1982relativistic, rybicki1991radiative}. Thus, particles that experience larger accelerations will experience more significant radiative losses. We note that radiation reaction plays a vastly different role from collisional effects in the plasma phase space, as the former constricts the phase-space volume while the latter expands it. Consequently, radiation reaction effectively reduces the entropy of the plasma particles, driving them away from thermodynamic equilibrium by increasing the entropy of the synchrotron photon spectrum. This will be demonstrated in this work. The phase-space cooling changes the plasma dynamics drastically, compared to the classical Lorentz Force, resulting in kinetically unstable distributions that can be a source of magnetic-field amplification or coherent radiation. We will demonstrate that this effect is relevant for isotropic Maxwellian (or Maxwell--J\"uttner) plasmas with a minimum temperature of $T>m_e c^2/\sqrt{3}$.

In this work, we focus on the synchrotron-cooling-dominated regime, i.e., when the effects of radiation reaction on a distribution of particles in a constant, strong magnetic field dominates the collective plasma dynamics, this occurs at large $\gamma$ and strong $B$. We emphasize that our work is in the context of the classical radiation reaction regime $\chi \ll 1$, where $\chi$ is the Lorentz- and Gauge-invariant parameter $\chi=e\sqrt{-(F_{\mu\nu} p^\nu)^2}/m_e^3$, $e$ is the elementary charge, $F_{\mu\nu}$ is the electromagnetic tensor, and $p^\nu$ is the four-momentum of the particle \cite{ritus1985quantum, di2012extremely}. For a constant background magnetic field, $\chi$ reduces to $\chi=p_\perp \left|\mathbf{B}\right| /(m_e B_\mathrm{Sc})$, where $B_{Sc} = m_e^2 c^2 /(e \hslash) \simeq 4.41\times10^9\, \mathrm{T}$ is the Schwinger critical field. Our results also qualitatively apply to the quantum regime $\chi > 1$, with the main difference being the momentum-diffusion effects from QED synchrotron emission, as demonstrated in preliminary simulations \cite{bilbao2022radiation}. QED radiation reaction can be accurately modeled by including a diffusive term in the Vlasov equation \cite{vranic2016quantum_m} (which is beyond the scope of this work).

This paper is organized as follows. In Sec. \ref{sec:rad_cool} we study the momentum-space trajectories of particles undergoing synchrotron cooling under the Landau--Lifshftz formulation. From the momentum trajectories of single synchrotron-cooled particles, we derive the conditions for a population inversion, i.e., momentum distribution function (MDF) $f (\mathbf{p},t)$ with regions that fulfill $\partial f/ \partial p_\perp > 0$, where $p_\perp$ is the momentum perpendicular to the magnetic field, i.e., ring-shaped MDFs, since our system is cylindrically symmetric along the magnetic field. This extends previous work\cite{bilbao2022radiation} from the synchrotron- to cyclotron-dominated regime, where radiation reaction does not give rise to momentum-space bunching, giving a natural condition on the development of population inversions. We then study the evolution of the MDF by including the radiation reaction force in the Vlasov equation \cite{kuz1978bogolyubov,hazeltine2004radiation, tamburini2011radiation, hirvijoki2015radiation, decker2016numerical}, from which general features of the evolving MDF, including the general ring-shaped pattern that arises during the evolution, and the relevant timescales, are derived.

In Sec. \ref{sec:tolerance}, we discuss competing processes, such as curvature or inhomogeneities in the guiding magnetic field, and collisional effects that might diffuse or inhibit the ring formation. Finally, we conclude in Sec. \ref{sec:conc}, that radiation reaction produces MDFs with inverted Landau populations, which provide free energy to drive instabilities and coherent radiation, such as the electron cyclotron maser instability \cite{le1984direct,bingham2000generation, cairns2005cyclotron, cairns2008cyclotron, melrose2016cyclotron}. We explore how these results are relevant for astrophysical and laboratory plasmas \cite{melrose1995models, lyutikov1999nature, davoine2018ion,melrose2021pulsar, lyutikov2021coherent}. We show that our findings apply to systems of particles undergoing formally equivalent cooling processes, such as beams interacting with laser pulses and particle beams in ion channels undergoing betatron cooling \cite{bilbaoinprpep2024}.

\section{Radiation reaction cooling}\label{sec:rad_cool}
We will consider the classical description of radiation reaction \cite{dirac1938classical, landau1975classical, hartemann1995classical, bell2008possibility, tamburini2010radiation}. The classical description of radiation reaction can be shown to be valid for $\chi\lesssim1$, as demonstrated in Appendix \ref{ap:qed}, where the effects of Quantum corrections are shown to be small compared to the classical prescription acting on a collection of synchrotron radiating particles. The radiation reaction force for an electron with arbitrary momentum in a constant electromagnetic field is described by the Landau--Lifshitz expression for radiation reaction \cite{landau1975classical, tamburini2010radiation}
\begin{multline}
\mathbf{F}_{RR} = -\frac{2}{3} \frac{e\alpha}{B_{\mathrm{Sc}} c^2} \left\{\frac{\gamma \mathbf{p}}{m_e c} \left[\left(\mathbf{E} + \frac{\mathbf{p}\times \mathbf{B}}{\gamma m_e c}\right)^2 -\left( \frac{\mathbf{p}\cdot\mathbf{E}}{\gamma m_e c}\right)^2\right]\right.\\
\left.-\mathbf{E}\times \mathbf{B} -\frac{\mathbf{B}\times \left( \mathbf{B}\times \mathbf{p}\right)+\mathbf{E}\left(\mathbf{p}\cdot\mathbf{E}\right)}{\gamma m_e c} \right\},\label{eq:LL}
\end{multline}
where $\alpha$ is the fine-structure constant, $\gamma = \sqrt{1 + \mathbf{p}^2/m_e^2c^2}$, and $\mathbf{E}$ and $\mathbf{B}$ are the electric and magnetic fields (in c.g.s. units). The first term in Eq. \eqref{eq:LL}, which dominates for relativistic particles ($\gamma \gg 1$), already shows a non-linear dependence of the radiation reaction force on the momentum of the particle $\mathbf{p}$. To study synchrotron cooling, we consider the case in which $\mathbf{E} = 0$ and constant magnetic field $\mathbf{B}=B\,\hat{e}_\parallel$, where $\hat{e}_\parallel$ is the unit vector along the magnetic-field direction. Thus Eq. \eqref{eq:LL} simplifies to
\begin{equation}
    \mathbf{F}_{RR} = -\frac{2}{3} \frac{e\alpha}{\gamma B_{\mathrm{Sc}} c ^{2}}\left[\frac{\mathbf{p}\left( \mathbf{p}\times \mathbf{B}\right)^2}{m_e^3 c^3}  -\frac{\mathbf{B}\times \left( \mathbf{B}\times \mathbf{p}\right)}{m_e c} \right].\label{eq:LLBfield}
\end{equation}

We now focus on the single-particle momentum evolution due to synchrotron cooling. Due to the symmetry perpendicular to the magnetic-field direction, it is convenient to decompose the momentum vector $\mathbf{p}$ into the parallel $p_\parallel$ and the perpendicular $p_\perp$ momentum components with respect to $\mathbf{B}$. Thus, the cross products in Eq. \eqref{eq:LLBfield} simplify to $\left(\mathbf{p}\times \mathbf{B}\right)^2 = p_\perp^{2} B^{2}$ and $\mathbf{B}\times\left( \mathbf{B}\times\mathbf{p}\right) = -B^{2} p_\perp{\hat{e}_\perp}$. From now on, momentum $\mathbf{p}$ and time $t$,  are given in units of $m_ec$, and the inverse of the cyclotron frequency $\omega_{ce}^{-1} = m_e c/eB$, respectively. And we define $B_0 = B/B_{\mathrm{Sc}}$.

\begin{figure}[t]
\includegraphics[width=\linewidth]{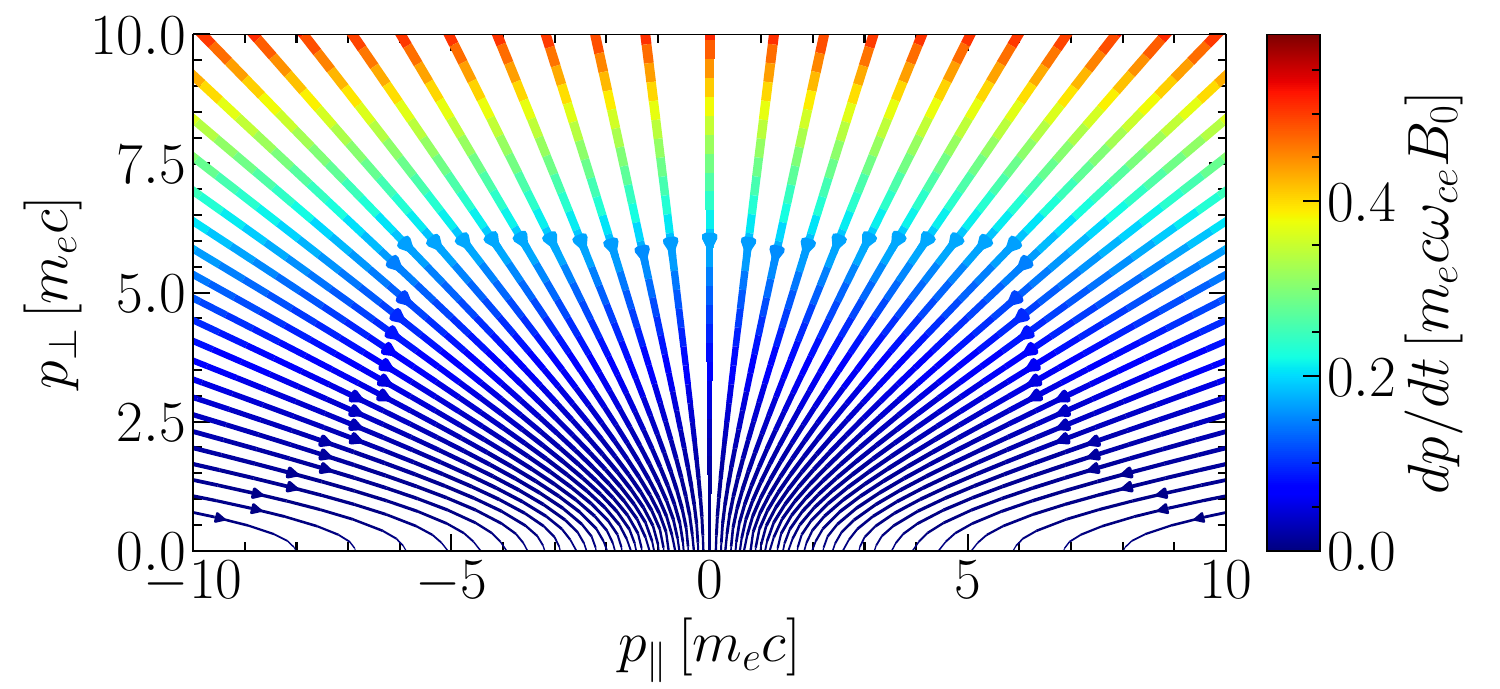}
    \caption{Streamlines of the momentum-space trajectories of particles undergoing synchrotron cooling [Eqs. \eqref{eq:perpt} and \eqref{eq:parat}] demonstrating that particles in a strong magnetic field cool down anisotropically and in a non-linear fashion. This is seen by the increase in the gradient of the magnitude of the radiation force as a function of $p_\perp$ indicated by the color of the streamlines, stronger (red) at larger $p_\perp$ than at lower $p_\perp$.}
    \label{fig:trajectories}
\end{figure}

The equations of motion due to synchrotron cooling are
\begin{eqnarray}
    \frac{d p_\perp}{d t} &=& -\frac{2}{3} \alpha B_0 \frac{p_\perp+p_\perp^3}{\gamma} ,\label{eq:LLBfield_perp}\\
    \frac{d p_{\parallel\ }}{d t} &=& -\frac{2}{3} \alpha B_0\frac{p_\parallel p_\perp^2 }{\gamma}.\label{eq:LLBfield_parallel}
\end{eqnarray}
One can show that, for any given initial perpendicular $p_\perp(t=0) = p_{\perp 0}$, parallel $p_\parallel(t=0) = p_{\parallel 0}$ momentum, and corresponding Lorentz factor $\gamma_0= \sqrt{1+p_{\parallel 0}^2+p_{\perp 0}^2}$,  the exact trajectory in momentum space follows
\begin{eqnarray}
   p_\perp(t) &=& \frac{p_{\perp 0}}{\cosh{(\tau')}\left[1+\sqrt{1+p_{\perp 0}^2} \tanh{(\tau')}\right]},\label{eq:perpt}\\
   p_\parallel(t)&=&p_{\parallel 0 } \frac{1+ \tanh{(\tau')}/\sqrt{1+p_{\perp 0}^2}}{1+ \tanh{(\tau')}\sqrt{1+p_{\perp 0}^2}}\label{eq:parat},
\end{eqnarray}
where $\tau' = \tau \sqrt{1+p_{\perp 0}^2}/\gamma_0$ and $\tau= 2\alpha B_0 t/3$. We observe that a particle with initial $p_{\perp 0} \to\infty$ follows the trajectory $p_\perp (t) = 1/\sinh(\tau)$, implying that the whole momentum space is bounded within that region and is compressed with time. The exact solution of the momentum-space trajectories, given by Eqs. \eqref{eq:perpt} and \eqref{eq:parat}, are shown in Fig. (\ref{fig:trajectories}), Figure (\ref{fig:trajectories}) shows how the radiative cooling increases in a non-linear manner with $p_\perp$. The color, which is proportional to the cooling rate, changes more rapidly at higher $p_\perp$ and remains mostly constant along a varying $p_\parallel$ and constant $p_\perp$ line. Note that as $t\to\infty$, the momentum trajectories converge towards $p_\perp \to 0$ and $p_\parallel \to p_{\parallel 0}/\sqrt{1+ p_{\perp0}^2}$. The latter is a result of the covariant nature of the Landau--Lifshitz formulation of radiation reaction, and is associated with one of the constants of motion,
given by 
\begin{equation}
    C_{1} = \frac{p_{\parallel}}{\sqrt{1+p_{\perp}^{2}}}\label{eq:c1}.
\end{equation}
$C_1$ characterises the streamlines shown in Fig. (\ref{fig:trajectories}). The second constant of motion can be obtained from Eq. (\ref{eq:perpt}) and Eq. (\ref{eq:parat})
\begin{equation}
    C_{2} = \tanh^{-1}\left(\frac{1}{\sqrt{1+p_{\perp}^{2}}}\right) - \frac{\gamma}{\sqrt{1+p_{\perp}^{2}}}\tau,\label{eq:c2}
\end{equation}
which is time-dependent, since radiation reaction is a dissipative process.

We now employ the single-particle trajectories to understand the collective effect of synchrotron cooling, particularly the conditions for an initial MDF $f_0 \equiv f(t=0)$ to develop a population inversion $\partial f/\partial p_\perp >0$ in a finite time. As the population-inversion region in momentum space occurs where $p_\perp\gg p_\parallel$ \cite{bilbao2022radiation}, we will focus our study on the evolution of the MDF in that region. There, the momentum-distribution evolution is dominated by the cooling in $p_\perp$.

We consider the evolution of a small volume of momentum space $V_p(t)$. As radiation reaction deforms this volume, it will always retain the same number of particles, implying
\begin{multline}
    \iint_{V_{p}(t)}\mathrm{d}p_{\parallel} \mathrm{d}p_{\perp}p_{\perp}f(p_{\perp},p_{\parallel},t) = \\\iint_{V_{p}(0)}\mathrm{d}p_{\parallel} \mathrm{d}p_{\perp}p_{\perp}f_{0}(p_{\perp},p_{\parallel}).
\end{multline}
By making the change of variables $\lbrace p_{\perp},p_{\parallel}\rbrace \to \lbrace{p_{\perp 0},p_{\parallel 0} \rbrace}$ in the left-hand side (analogous to the method of characteristics) we may compare the integrands, finding
\begin{equation}
    f(p_\perp,p_{\parallel},t) = \left|\frac{\partial (p_\perp,p_{\parallel})}{\partial (p_{\perp0},p_{\parallel0})}\right|^{-1}\frac{p_{\perp 0}}{p_{\perp }} f_0(p_{\perp 0},p_{\parallel 0}).\label{eq:evof_reg}
\end{equation}
Equation \eqref{eq:evof_reg} simplifies considerably at $p_{\parallel}=0$ (the region where we expect the population inversion to take place\cite{bilbao2022radiation}). By taking the derivative of this with respect to $p_{\perp 0}$ and  using Eq. \eqref{eq:perpt}, we can find the regions that eventually develop $\partial f/\partial p_\perp>0$ fulfill
\begin{multline}
    \frac{1}{p_\perp f_0(p_\perp)} \frac{\partial f_0(p_\perp,t)}{\partial p_\perp} >\\ -\frac{2  \tanh (\tau) \left(2 \gamma_\perp^2+3 \gamma_\perp  \tanh (\tau)+1\right)}{\gamma_\perp ^2
   (\gamma_\perp +\tanh (\tau)) (\gamma_\perp  \tanh (\tau)+1)},
\end{multline}
where $\gamma_\perp = \sqrt{1+p_\perp^2}$ and we have dropped the $0$ subscript as only initial-condition parameters remain. At late times ($\tau\to\infty$) $\tanh(\tau)\to 1$. Thus, the condition for a distribution function to develop inverted Landau populations is
\begin{equation}
     \frac{1}{p_\perp f_0(p_\perp
     )} \frac{\partial f_0(p_\perp,t)}{\partial p_\perp} > - \frac{2(1+2\gamma_\perp)}{\gamma_\perp^2 (1+\gamma_\perp)}.\label{eq:condition_instability}
\end{equation}

For a Maxwellian MDF, $\partial f_{0}/\partial p_\perp = -p_\perp f_{0}/p_{\mathrm{th}}^2$, where $p_{\mathrm{th}}=\sqrt{T}$ and $T$ is the plasma temperature in units of $m_ec^2$. Thus, ring-shaped MDF will form when $p_{\mathrm{th}} > m_e c/\sqrt{3}\approx 0.57\, m_e c$, meaning that there is a minimal temperature of $T> \SI{295}{\kilo\electronvolt} \sim \SI{3e9}{\kelvin}$ required for the onset of these MDFs. For a Maxwell--J\"uttner distributions, $f_0 \propto e^{-\gamma/p_{\mathrm{th}}}$, one finds that $p_{\mathrm{th}} > m_e c/3\approx 0.33\, m_e c$  fulfills Eq. \eqref{eq:condition_instability}.

These examples demonstrate that MDFs with inverted Landau populations are a result of Vlasov--Maxwell's dynamics in the presence of radiation reaction for relativistic thermal plasmas. The inequality in Eq. \eqref{eq:condition_instability} is fulfilled by the most common distribution functions with sufficient thermal energy, such as Maxwell--Boltzmann, Maxwell--J\"uttner, power-laws, etc. Numerical tests have been performed with relativistic particle pushers and full Particle-in-Cell simulations with the OSIRIS code \cite{fonseca2002osiris, vranic2016classical_m, vranic2016quantum_m}, which corroborate our findings.

Generalized kinetic equations for non-conservative forces, particularly for radiation reaction, have been known for many decades \cite{hakim1968relativistic, hakim1971collective, kuz1978bogolyubov}. Recent results have employed generalized kinetic equations with radiation reaction to model conditions for experimental fusion \cite{hazeltine2004radiation, hirvijoki2015radiation, decker2015bump,decker2016numerical, stahl2015effective}, laser-plasma interactions \cite{tamburini2011radiation, berezhiani2008plasma}, and to derive fluid descriptions that include radiation reaction effects \cite{hazeltine2004closed}. Here, we employ the non-manifestly covariant form of the Vlasov equation with radiation reaction force term \cite{kuz1978bogolyubov, hazeltine2004radiation, tamburini2011radiation,hazeltine2004radiation, hirvijoki2015radiation, decker2015bump,decker2016numerical, stahl2015effective}
\begin{equation}
    \frac{\partial f}{\partial t}+ \frac{\mathbf{p}}{\gamma} \cdot \mathbf{\nabla}_r f+ \mathbf{\nabla}_p \cdot \left[ \left( \textbf{F}_{RR}+ \textbf{F}_{L} \right) f \right] = 0, \label{eq:pre_Vlasov}
\end{equation}
where $f (\mathbf{p},\mathbf{r},t)$ is the distribution function, $\mathbf{F}_{L}$ is the Lorentz force, and $\mathbf{\nabla}_{r}$ and $\mathbf{\nabla}_{p}$ are nabla operators acting on the position and momentum coordinates, respectively. Including the radiation reaction force as the operator $\mathbf{\nabla}_p \cdot \left( \mathbf{F}_{RR} f \right)$ guarantees the conservation of the number of particles \cite{stahl2015effective}. Since $\mathbf{F}_{L}$ conserves the phase-space volume, but $\mathbf{F}_{RR}$ is dissipative, then $\mathbf{\nabla}_p \cdot \mathbf{F}_{rad} \ne\mathbf{\nabla}_p \cdot \mathbf{F}_{L} = 0$. Since we consider a spatially homogeneous plasma, we can neglect the term proportional to $\mathbf{\nabla}_{r} f$. Moreover, as we are assuming cylindrical symmetry $f = f(p_\perp, p_\parallel)$, the effect of the Lorentz force due to a strong magnetic field on the distribution is $\mathbf{\nabla}_p \cdot \left(  \mathbf{F}_{L} f \right) = 0$, even when $\mathbf{F}_{L} \ne 0$. Thus, Eq. \eqref{eq:pre_Vlasov} simplifies to
\begin{equation}
    \frac{\partial f}{\partial t}+ \mathbf{F}_{RR}\cdot \mathbf{\nabla}_p f + f \mathbf{\nabla}_p \cdot\mathbf{F}_{RR} = 0. \label{eq:modVlasov_reduced}
\end{equation}
In cylindrical coordinates, the operators are $\mathbf{\nabla}_p f = \partial f/\partial p_\parallel \hat{\mathbf{e}}_\parallel + \partial f/\partial p_\perp \hat{\mathbf{e}}_\perp$ and 
\begin{eqnarray}
    \mathbf{\nabla}_p \cdot\textbf{F}_{RR} &=& \frac{1}{p_\perp} \frac{\partial F_{RR\perp}}{\partial p_\perp} + \frac{\partial F_{RR\parallel}}{\partial p_\parallel}\nonumber\\  
    \mathbf{\nabla}_p \cdot\textbf{F}_{RR} &=&-\frac{2}{3} \alpha B_0 \frac{2+4p_\perp^2}{\gamma}.\label{eq:div}
\end{eqnarray}
One notes that, as expected, the divergence of the radiation reaction force is negative. This demonstrates one of the key features of radiation reaction, i.e., radiation reaction compresses the momentum phase--space volume as $\mathbf{\nabla}_p \cdot\textbf{F}_{RR}<0$. In doing so, it acts against collisions that have the property $\mathbf{\nabla}_p \cdot\textbf{F}_{col}>0$, where $\textbf{F}_{col}$ is an effective collisional force. Therefore, radiation reaction reduces the phase-space volume and thus, the entropy of the particles, decreases.

The Vlasov equation for a distribution function $f$ undergoing synchrotron cooling is
\begin{equation}
    \frac{\partial f}{\partial \tau}- \frac{p_\perp^3 + p_\perp}{\gamma} \frac{\partial f}{\partial p_\perp} - \frac{p_\perp^2 p_\parallel}{\gamma} \frac{\partial f}{\partial p_\parallel} - \frac{2+4p_\perp^2}{\gamma} f = 0. \label{eq:vlasov_eq}
\end{equation}
Equation \eqref{eq:vlasov_eq} can be numerically integrated for any given initial distribution $f_0$. This differential equation describes the non-linear transport in momentum space where the momentum-space flow is compressible, which can exhibit momentum-space shocks analogous to hydrodynamic shocks \cite{bilbao2022radiation,zhdankin2022synchrotron}, resulting from $\mathbf{\nabla}_p \cdot\textbf{F}_{RR}<0$ and $\mathbf{\nabla}_p \cdot\textbf{F}_{RR}$ not being constant along $p_\perp$. 

Our key findings apply beyond synchrotron-cooled plasmas. Other systems or electromagnetic-field configurations where the dissipative power depends non-linearly on the energy level occupied will develop population inversions. In our current case of synchrotron-cooled plasmas, the radiative power $P \propto p_\perp^3 /\gamma$ depends non-linearly on the Landau energy level occupied (i.e., $p_\perp$). Other radiation cooling mechanisms that exhibit analogous behavior, such as is the case of electrons undergoing betatron motion in an ion-channel, whose radiative power $P \propto r_\beta^2$, depends non-linearly on the betatron energy level occupied, i.e., the betatron oscillation amplitude $r_\beta$, and will be the subject of future work \cite{bilbaoinprpep2024}.

\begin{figure}
    \includegraphics[width=\linewidth]{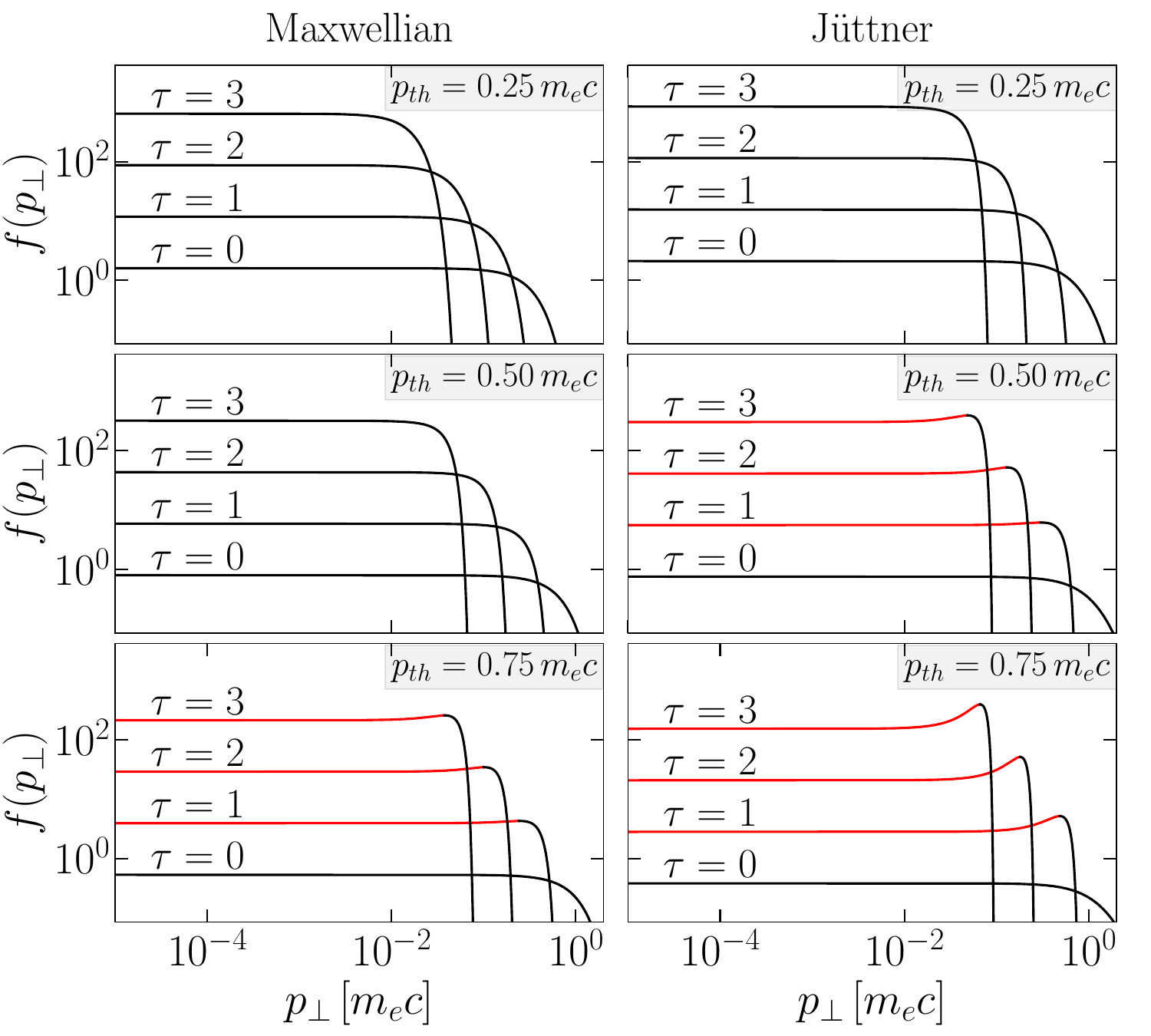}
    \caption{The time evolution of Eq. \eqref{eq:distribution_full} for an initial Maxwellian (left) and Maxwell--J\"uttner plasma (right) demonstrating the minimum thermal energy needed to develop a region with $\partial f/\partial p_{\perp}>0$, indicated in red. Different panels show different initial thermal spreads, $p_{\mathrm{th}}=0.25\, m_e c$ (top), $p_{\mathrm{th}}=0.50\, m_e c$ (middle), and $p_{\mathrm{th}}=0.75\, m_e c$ (bottom), at different normalized times $\tau =0,\,1,\,2\,\text{, and}\,3$, in unnormalized units $t= 2m_e B_\mathrm{Sc}\tau/(3\alpha e B^2)$.}
    \label{fig:dist_evo}
\end{figure}
We note that analytical solutions to Eq. \eqref{eq:vlasov_eq} exist. For example, one can consider a distribution function with a large spread in $p_\perp$ centered around $p_\parallel=0$ and an infinitesimal width in $p_\parallel$. In this configuration, one can approximate $\tau'=\tau$ (since $p_\perp\gg p_\parallel$) and employ the method of characteristics to solve Eq. \eqref{eq:vlasov_eq} for $f(p_\perp, p_\parallel =0) = f_\perp (p_\perp)$, finding
\begin{equation}
    f_\perp(p_\perp, t)=\frac{f_{\perp0}\left(\text{csch}\left(a\right)\right)}{\left(\gamma_\perp p_\perp\mathrm{sinh}(a)\mathrm{tanh}(a)\right)^2}, \label{eq:distribution_full}
\end{equation}
where $a=\log \left(p_{\perp} /(\gamma_{\perp}+1)\right)+\tau$ and $\gamma_{\perp} = \sqrt{1+p_\perp^2}$. Equation (\ref{eq:distribution_full}) has a singularity at $p_\perp = 1/\sinh(\tau)$. This singularity lies outside the range of valid physical values for $p_\perp$, as noted before, the whole momentum space is bounded within $p_\perp<1/\sinh{(\tau)}$; therefore Eq. (\ref{eq:distribution_full}) is well-behaved for the range of values $0<p_\perp<1/\sinh{(\tau)}$.

The results from plotting Eq. \eqref{eq:distribution_full} for different initial distributions (Maxwellian and Maxwell--J\"uttner) functions are shown in Fig. (\ref{fig:dist_evo}). Regions where the curves are red in Fig. \ref{fig:dist_evo} have $\partial f/\partial p_\perp >0$. These results validate our earlier results that Maxwellian and Maxwell--J\"uttner distributions necessitate $p_{\mathrm{th}}>0.57\, m_ec$ and $p_{\mathrm{th}}>0.33\, m_ec$, respectively, to develop ring-shaped MDFs. As we have shown that plasmas need a minimum thermal energy to develop a population inversion, we will henceforth study Eq. \eqref{eq:vlasov_eq} in the relativistic regime, that is also the regime relevant for astrophysical plasmas.

When $\gamma\gg1$ and $p_\perp\gg1$, the trajectories in momentum space [Eqs. \eqref{eq:perpt} and \eqref{eq:parat}] and the Vlasov equation [Eq. \eqref{eq:vlasov_eq}]  become
\begin{gather}
   p_\perp(t) = \frac{p_{\perp 0}}{1+p_{\perp 0}^2 \tau/ \gamma_0},\label{eq:perpt_rel}\\
   p_\parallel(t)= \frac{p_{\parallel 0 }}{1+p_{\perp 0}^2 \tau/ \gamma_0}\label{eq:parat_rel},\\
       \frac{3}{2\alpha B_0} \frac{\partial f}{\partial t}- \frac{p_\perp^3}{\gamma} \frac{\partial f}{\partial p_\perp} - \frac{p_\perp^2 p_\parallel}{\gamma} \frac{\partial f}{\partial p_\parallel} - \frac{4p_\perp^2}{\gamma} f = 0. \label{eq:vlasov_eq_rel}
\end{gather}
Equation (\ref{eq:vlasov_eq_rel}) can be solved if one examines regions of momentum space where $p_\perp \sim \gamma$ (i.e., $ p_\perp \gg p_\parallel$). Then, a solution for Eq. (\ref{eq:vlasov_eq_rel}) can be obtained using method of characteristics \cite{bilbao2022radiation}
\begin{equation}
    f(p_\perp, p_\parallel, t) = \frac{f_0\left(\frac{p_\perp}{1- \tau p_\perp}, \frac{p_\parallel}{1- \tau p_\perp }\right)}{\left( 1- \tau p_\perp \right)^4}. \label{eq:general_solution}
\end{equation}
Equation (\ref{eq:general_solution}) describes the evolution of an initial momentum distribution function $f_0$ due to synchrotron cooling in the relativistic limit. Similarly to Eq. (\ref{eq:distribution_full}), Eq. (\ref{eq:general_solution}) appears to have a singularity at $p_\perp = 1/\tau$. It can be similarly shown that such point always lies outside the range of validity of Eq. (\ref{eq:general_solution}). From Eq. (\ref{eq:perpt_rel}), a particle with initial $p_{\perp0} \to \infty$ evolves following $p_\perp(t) = 1/\tau$; thus all particles are bounded between $0<p_\perp<1/\tau$. This means that $f$ is well-behaved within the range of values of $p_\perp$ that are physically relevant, \emph{i.e.} $p_\perp < 1/\tau$.

\begin{figure}
    \includegraphics[width=\linewidth]{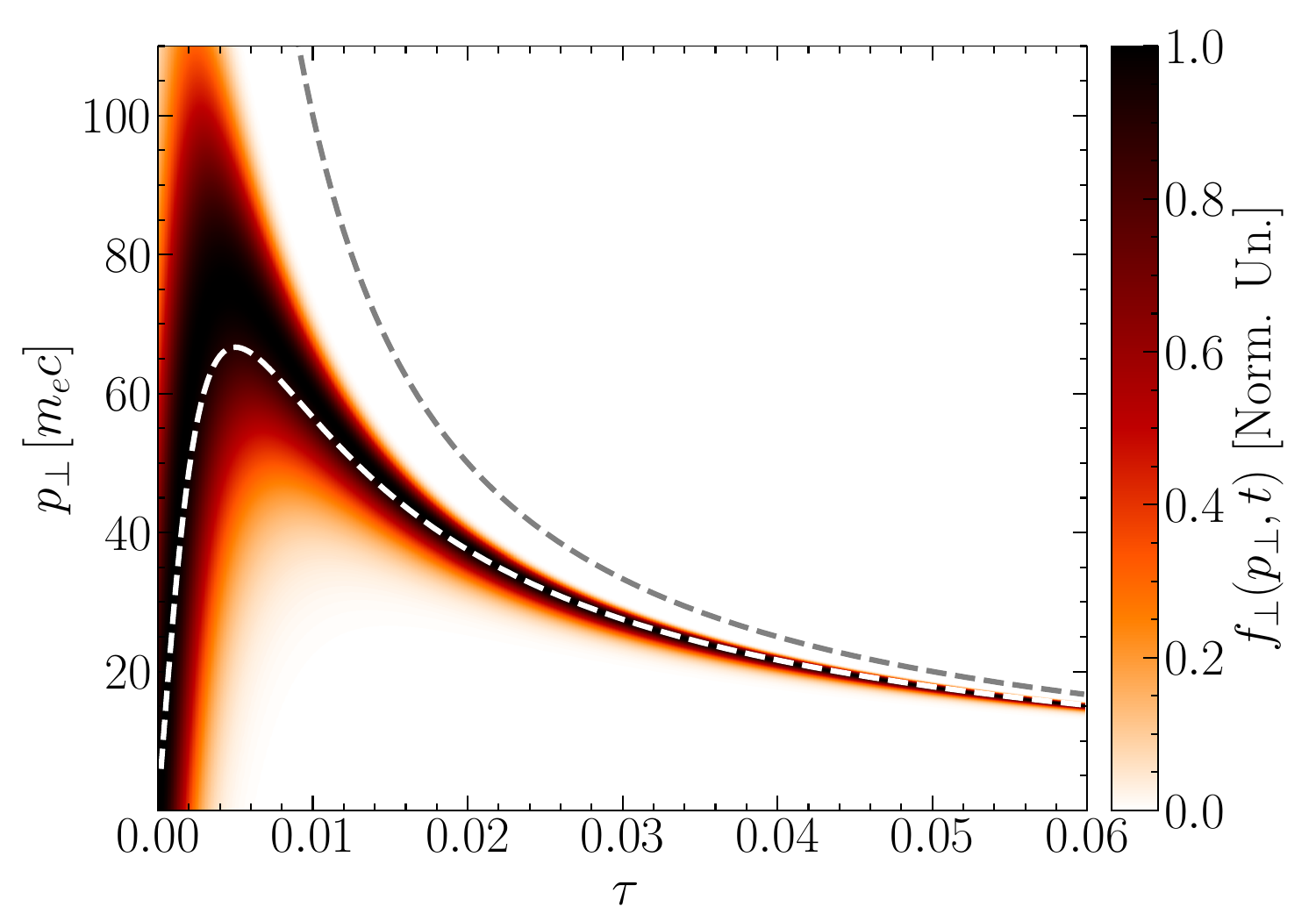}
    \caption{Evolution of an isotropic initial Maxwellian distribution, with $p_{\mathrm{th}}=100\, m_e c$ calculated analytically from Eq. \eqref{eq:general_solution} (The distribution is normalised to its maximum value at each time). The white and gray dashed lines indicate the ring radius predicted by Eq. \eqref{eq:peakpos} and the curve that bounds the momentum space $p_{\perp}=\tau^{-1}$, respectively.}
    \label{fig:ring_radius_evo}
\end{figure}

As we showed before, relativistic plasmas undergoing synchrotron cooling will develop a population inversion $\partial f/\partial p_\perp >0$. To determine the relevant timescales for forming these ring-shaped MDFs one can employ a representative initial distribution function with a large spread in $p_\perp$. We consider an isotropic Maxwellian distribution function 
\begin{equation*}
    f_{0,\text{MB}}(p_\perp, p_\parallel)=\frac{1}{(2\pi p_{\mathrm{th}}^2)^{3/2}}\exp\left(-\frac{p_\perp^2 + p_\parallel^2}{2p_{\mathrm{th}}^2}\right).
\end{equation*}
The resulting ring radius $p_R(t)$ in momentum space, defined as $\left|\partial_{p_\perp} f_\perp(p_\perp, t)\right|_{p_\perp =p_R(t)} = 0$, where $f_\perp(p_\perp, t) = \int_{-\infty}^{\infty} f(p_\perp, p_\parallel, t ) d p_\parallel$ is the integrated distribution along $p_\parallel$. Using Eq. (\ref{eq:general_solution}) to determine the temporal evolution of $f_{0,\text{MB}}$, $p_R$ evolves as
\begin{equation}
    p_R (t) = \frac{1+ 6 p_{\mathrm{th}}^2 \tau^2 - \sqrt{1+12 p_{\mathrm{th}}^2 \tau^2}}{6 p_{\mathrm{th}}^2\tau^3}. \label{eq:peakpos}
\end{equation}
Several conclusions about the formation and evolution of ring momentum distributions can be drawn from Eq. (\ref{eq:peakpos}). Figure (\ref{fig:ring_radius_evo}) shows the evolution of $f_\perp(p_\perp,t)$ for an initial isotropic Maxwellian plasma with $p_{\mathrm{th}}=100\, m_e c$ and the ring radius evolution (white dashed line) according to Eq. \eqref{eq:peakpos}. At early times $\tau=2\alpha B_0 t/3\ll 1$, the ring radius grows linearly with time as $p_R(t)\sim 2 p_{\mathrm{th}}^2 \alpha B_0 t$; this is due to a build-up of particles in momentum space at lower $p_\perp$ first. The ring MDF results from bunching as particles with higher energies radiate strongly and slow down faster, catching up with the lower energy particles. The bunching begins at low $p_\perp$ and propagates towards higher $p_\perp$ values, eventually, it asymptotically approaches the boundary of momentum space $p_\perp<p_\perp^*=\tau^{-1}$ and $p_R(t)\sim 1/\tau=3/(2\alpha B_0 t)$, for large $\tau$ (in Fig. \ref{fig:ring_radius_evo} at $\tau\sim0.05$).

Whenever the bunching direction reverses [maximum value of $p_R (t)$], the gradient of the MDF starts growing much faster, and the distribution will produce a higher contrast ring-shaped MDF. Thus, we define that time as the ring formation time normalized to cyclotron periods
estimated to be 
\begin{equation}
    t_R = \frac{3}{4 \alpha B_0 p_{\mathrm{th}}} = \frac{3}{4 \alpha \chi_{\mathrm{th}}}, \label{eq:ring_formtime}
\end{equation}
where $\chi_{\mathrm{th}} = p_{\mathrm{th}} B_0$, in unnormalized units $t_R= 3 m_e^2 c^2 B_\mathrm{Sc}/(4\alpha e B^2 p_\mathrm{th})$. 

A second longer timescale can be estimated, related to the ring MDF evolution to achieve efficient conditions for the growth of kinetic instabilities, specifically the electron cyclotron maser instability. There are two relevant factors here: i) the generation of population inversion $\partial f/\partial p_\perp$ with high gradients and ii) relativistic inertial effects that slow down the instability growth.
We can estimate the evolution of $\partial f/\partial p_\perp$ by noting there is a small region between the momentum-space boundary, given by $p_\perp = 1/\tau$ [gray dashed curve in Fig. \ref{fig:ring_radius_evo}], and the ring radius [white dashed curve in Fig. \ref{fig:ring_radius_evo}] of size $\delta p = 1/\tau - p_R(t)$, which can be approximated, from Eq. \eqref{eq:peakpos}, for initially large $p_{\mathrm{th}}$ as $\delta p = 1/(6 p_{\mathrm{th}} \tau)$. Within that region, the distribution function has the property $\partial f / \partial p_\perp <0$; conversely, there is a region just below the ring radius of comparable width $\delta p$, where the distribution function has the property $\partial f / \partial p_\perp >0$. Thus, the positive gradient can be approximated as $\partial f /\partial p_\perp \sim f(p_R)/\delta p$ and $f(p_R) \sim  1/4\pi p_R(t) \delta p$, leading to
\begin{equation}
    \frac{\partial f}{\partial p_\perp} \sim \frac{9}{\pi} \frac{p_{\mathrm{th}}^2}{p_R(t)^5} = \frac{9}{\pi} p_{\mathrm{th}}^2 \tau^5,
\end{equation}
which shows how the gradient, at early times $\tau\ll1$, increases slowly as a function of $\tau$ and, as $\tau\lesssim1$, the gradient grows faster. For a fixed gradient ring MDF, the maser growth rate is maximum when $p_R \sim 1$; for $p_R \gg 1$, relativistic inertial effects decrease the growth rate \cite{thalesinprpep2024}. Thus, a natural choice for the maximum maser onset timescale in cyclotron periods is $t_i$ the time it takes for $p_R \sim 1$, which for $p_{\mathrm{th}}\gg 1$ occurs at $\tau = 1$, resulting in
\begin{equation}
    t_i = \frac{3}{2\alpha B_0},\label{eq:ring_evotime}
\end{equation}
in unnormalized units $t_i = 3 m_e c B_\mathrm{Sc}/(2\alpha e B^2)$. Within the timescale $t_i$, the growth rate associated with the maser instability has developed a high gradient $\partial f / \partial p_\perp \gg 1$, and the ring has cooled down enough that relativistic inertial effects have been reduced. Thus, the onset of the maser instability should occur within a shorter timescale than $t_i$.

Scaling engineering formulas can be obtained for both timescales $t_R$ and $t_i$, 
\begin{eqnarray}
    t_R\, \left[\SI{5.2}{\nano \second}\right] &\simeq& B^{-2}\, \left[\SI{50}{\mega\gauss}\right] T_e^{-1}\, [\SI{10}{\mega e \volt}],\label{eq:engR} \\
t_i\, \left[\SI{205}{\nano\second}\right]&\simeq& B^{-2}\, \left[\SI{50}{\mega \gauss}\right].\label{eq:engI}
\end{eqnarray}
These scalings are compatible with astrophysical conditions provided by the magnetosphere of compact objects, which can greatly exceed $B\gg\SI{100}{\mega\gauss}$.

\section{Effects of curved and inhomogeneous magnetic fields, and collisional effects}\label{sec:tolerance}
So far, our study has focused on the onset and evolution of ring-shaped MDFs within ideal magnetic-field configurations. To understand the regimes in which these ring distributions can emerge from synchrotron cooling, we will assess the validity of the presented model and determine the resilience of the cooling mechanism against other processes that may diffuse or alter the evolution of the radiative cooling process. We will discuss the effects of curved magnetic-field configurations; the impact of inhomogeneous magnetic fields, such as mirror fields or compressional Alfv\'en waves; and the diffusive effects of collisions.

Curvature effects can be relevant in a macro-scale, such as the case of a beam propagating along a curved magnetic field. As a result, beam particles will experience curvature and other drifts. A first estimate for the validity of our prediction for generating kinetically unstable MDFs [Eq. \eqref{eq:engI}] holds as long as that timescale is much shorter than the light crossing time of the curvature radius ($r_c$), leading to 
\begin{equation}
r_c \gg \frac{3 c}{2\alpha B_0 \omega_{ce}} = \frac{3 m_e c }{2 \alpha e } \frac{B_\mathrm{Sc}}{B^2}. \label{eq:radius}
\end{equation}
This scaling can be written as $r_c\, [\SI{}{\kilo\metre}] \gg 1.5 /(B\, [\SI{10}{\mega\gauss}])^{2}$, which is compatible with astrophysical compact objects whose scale exceeds kilometers and whose magnetic fields greatly exceed $\SI{10}{\mega\gauss}$. In the regime where Eq. \eqref{eq:radius} is not fulfilled, synchrotron cooling will still efficiently occur, and curvature may modify the distribution function. Our preliminary results of beams undergoing synchrotron cooling in curved magnetic-field lines show that including curvature drifts leads to a spiral within a constant pitch-angle in momentum space \cite{assunçao2024}, which is also an unstable distribution.

The scaling provided by Eq. \eqref{eq:radius} hints that synchrotron cooling might not be easily tested under current laboratory setups. Nonetheless, configurations such as betatron cooling may provide easier access to probing the properties of radiatively cooled plasmas with current technology.

At scales smaller than the curvature radius, inhomogeneities of the guiding magnetic field can arise from magnetic turbulence or the propagation of compressional Alfv\'en waves, providing mirror fields that can scatter the ring-shaped MDF and diffuse it. If the mirror interaction occurs within a timescale shorter than the ring evolution $t_i$, we can assume that the magnetic moment $\mu = p_\perp/B^2$ and the Lorentz factor $\gamma$ are constant during the interaction. The ring MDF will be trapped by the mirror and scattered when its pitch angle is greater than the critical angle of the mirror given by $\sin\,\theta_c = \sqrt{B_0/B_A}$, where $B_0$ is the guiding field and $B_A$ is the peak magnetic-field strength in the mirror field \cite{nicholson1983introduction}.

From the insights obtained from studying radiatively cooled thermal plasmas analytically and numerically \cite{bilbao2022radiation}, we know that particles in beams with a given perpendicular momentum spread $\Delta p_\perp$ and average Lorentz factor $\gamma_b$ cool down towards $p_\perp = p_\parallel \simeq 0$ and cool down faster in $p_\perp$ as seen by the trajectories in Fig. \ref{fig:trajectories}. Therefore, the resulting ring-beam MDF has a ring radius smaller than $\Delta p_\perp$ and an average $\gamma$ lower than $\gamma_b$. This results in a beam pitch angle $\theta < \Delta p_\perp/\gamma_b$.

For such a beam propagating along a guiding field of strength $B_0$ to be scattered by a magnetic mirror of strength $B_A$, the beam pitch angle must be smaller than the critical mirror angle. Thus, 
\begin{equation}
    \frac{B_0}{B_A} < \sin^{2}\left( \frac{\Delta p_\perp}{\gamma_b}\right).
\end{equation}
For small divergences we take $\sin(\theta) \sim \theta$ and obtain that $B_A \gtrsim B_0 /\theta^{2}$ for the ring to be diffused. For the case of astrophysical beams with milirad divergences \cite{miniati2013relaxation} we estimate that $B_A \gtrsim 10^6 B_0$, such gradient is not easily achievable via magneto turbulence or by compressional Alfv\'en waves when the guiding field is on the order of a gigagauss, as expected around compact objects. Thus, in astrophysical scenarios, we conclude that synchrotron-induced ring beams are resilient to the interaction with a turbulent medium or compressional Alfv\'en waves.

We have assumed a regime where the synchrotron cooling timescales and the resulting plasma physics occur in a timescale where collisional relaxation cannot return the plasma to kinetic equilibrium, i.e., the collisionless regime. These results are valid for large magnetic fields; however, if one considers the small magnetic-field limit $B\to 0$, then $t_R,\, t_i\to \infty$, according to Eqs. \eqref{eq:ring_formtime} and \ref{eq:ring_evotime}. In this regime, collisional effects could inhibit the ring formation and evolution. We now compare the ring evolution timescale $t_i$ to the relaxation timescales given by collisional processes. We consider three collisional processes capable of diffusing the ring momentum distribution: lepton-lepton $e^\pm+ e^\pm \to e^\pm+e^\pm$, lepton-ion $e^\pm+i \to e^\pm+i$, and Compton $e^\pm + \gamma \to e^\pm+ \gamma$ collisions. We note that pair production/annihilation $e^- + e^+ \rightleftharpoons \gamma + \gamma$ processes can also be relevant. However, unlike collisional processes, these will produce/evaporate the pair plasma from/to a photon gas and not necessarily destroy the ring. This will be investigated elsewhere.

For the case of lepton-lepton and lepton-ion collisions, one can employ the standard kinetic theory and obtain the relaxation timescales for lepton-lepton and lepton-ion collisions\cite{trubnikov1965particle, sivukhin1966coulomb, goldston1995introduction}
\begin{align}
    \tau_{ee} &=\frac{12 \pi^{3/2}}{\sqrt{2}}\frac{\epsilon_0^2 m_e^2 c^3}{e^4} \frac{ 1}{n_e \ln{\Lambda}},\\
    \tau_{ei} &= \frac{12 \pi^{3/2}}{\sqrt{2}}\frac{\epsilon_0^2 m_e^2 c^3}{e^4}  \frac{1}{n_i Z^2 \ln{\Lambda}}.
\end{align}
where $n_e$ is the lepton density, $n_i$ is the ion density, $\ln{\Lambda}$ is the Coulomb logarithm, and, as we are dealing with relativistic plasmas, we have approximated $v_e \sim c$. Comparing both relaxation times, assuming the fastest case of $Z=1$ against the kinetic instabilities timescale $t_i$, we obtain $t_i/\tau_{ee}\,  = 10^{-6} B^{-2}\, [\SI{1}{\gauss}]\,n_e\, [\SI{}{\centi\metre^{-3}}] \ln{\Lambda}$. For astrophysical conditions, i.e., magnetic fields on the order of a gigagauss, densities in the order of $n_e \sim 10^{24}\SI{}{\centi\metre^{-3}}$ are required for collisions to be comparable to the ring evolution time, and collisional effects should not disrupt the generation of ring MDFs. Conversely, for current laboratory magnetic-field strengths and plasma densities in the order of $B\sim \SI{10}{\mega\gauss}$ and $n_e\sim 10^{20}\SI{}{\centi\metre^{-3}}$ then $t_i/\tau_{ee} \sim 1 \ln{\Lambda}$, which hints that laboratory settings relying on strong magnetic fields might not be able to efficiently produce ring MDFs due to collisional effects.

The collision rate for Compton scattering is estimated as $\nu_{e\gamma} = 2 \sigma c n_\gamma$ \cite{thomson1906conduction, compton1923quantum}, where $\sigma$ is the Klein--Nishina cross-section and $n_\gamma$ is the photon density. In the high energy limit, and in the electron frame, the Klein--Nishina cross-section has a peak for forward collisions, so the cross-section can be approximated as $\sigma = \pi r_e^2 /\epsilon_\gamma'$ \cite{klein1929streuung}, where $r_e$ is the classical electron radius and $\epsilon_\gamma'$ is the photon energy in the electron frame in units of $m_e c^2$. For frontal collisions in the beam frame, the cross-section transforms to $\sigma = \pi r_e^2 /(\gamma_e \epsilon_\gamma)$, where $\gamma_e$ is the beam Lorentz factor and $\epsilon$ is now the photon energy in the lab frame in units of $m_e c^2$. Hence, the relaxation time can be approximated 
\begin{equation}
    \tau_{e \gamma} = \frac{4\hbar}{3c \sigma_T} \frac{ \left<\omega_\gamma\right> \left<\gamma_e\right>}{ n_\gamma },
\end{equation}
where $\sigma_T$ is the Thompson cross-section, $\left<\gamma_e\right>$ is the average Lorentz factor of the leptons, and we have taken $\epsilon_\gamma = \hbar \left<\omega_\gamma\right>$, where $\left<\omega_\gamma\right>$ is the average frequency of the photon gas, and $\hbar$ is the reduced plank constant. Assuming a blackbody spectrum for the photons, we can obtain $\hbar\left<\omega_\gamma\right>/n_\gamma = \pi^6 c^3 \hbar^3/(60 \zeta^2(3) k_B^2 T_\gamma^2)$  \cite{lightman1982relativistic}, and therefore $t_i/\tau_{e\gamma} = 4.6 B^{-2}\, [\SI{}{\gauss}]\, T_\gamma^2\, [\SI{}{e\volt}] \left<\gamma_e\right>^{-1}$. We find that $t_i \ll \tau_{e\gamma}$ for gigagauss field strengths and photon temperatures in the X-ray ($T_\gamma \sim \SI{}{\kilo e\volt}$), compatible with astrophysical plasmas \cite{lyutikov2012very}.

\section{Conclusions}
\label{sec:conc}
In this work, we have presented the process under which an initially kinetically stable plasma undergoing synchrotron cooling will develop into a momentum ring distribution. We have demonstrated that the cooling process is anisotropic, and one must consider that plasmas undergoing synchrotron cooling will be characterized by transverse momentum distributions with inverted Landau populations, i.e., a ring momentum distribution or non-monotonic pitch-angle beam distribution. 

The resulting ring momentum distributions are kinetically unstable and will drive kinetic instabilities; so far two distinct kinetically unstable regimes have been identified: when $\beta \gg 1$, where $\beta = 2\mu_0 \omega_{pe}^2 T_e/B^2$ is the plasma pressure to magnetic-field pressure ratio, the pressure anisotropy due to the anisotropic synchrotron cooling will lead to the firehose instability\cite{zhdankin2022synchrotron}, and, in contrast, when $\beta\ll1$ due to $\omega_{pe}\ll\omega_{ce}$ the inverted Landau population will dominate and lead to electron cyclotron maser emission\cite{bilbao2022radiation}. Nevertheless, preliminary results, which will be the subject of future work, have shown that for large $\beta$ and $\omega_{pe}\ll\omega_{ce}$ (relevant for low-density plasmas with initially relativistic thermal energies) the plasma transitions from $\beta\gg1$ to $\beta\ll1$, where the synchrotron firehose will be triggered, followed by the electron cyclotron maser.

The model presented in this work employed the classical formulation of radiation reaction from the Landau--Lifshiftz model. Future studies shall address the development of rings in the strong QED regime and how this description introduces a diffusive effect. Particle-in-cell simulations have shown that the LL model accurately predicts the ring radius evolution in the $\chi\sim 1$ regime \cite{bilbao2022radiation}. Moreover, as the plasma cools down $\chi\to0$, it transitions from QED synchrotron cooling to classical synchrotron cooling. A QED model will allow the study of the interaction between the synchrotron photons and the plasma via Compton scattering and the production of cascades or avalanches \cite{wadiasingh2019repeating}, where a single photon or lepton could self-generate the whole plasma and produce a ring distribution.

We have studied the timescales for the onset of ring distributions and subsequent kinetic instabilities, from which we have concluded that the ring momentum distributions under the astrophysical conditions provided by compact objects must be pervasive, resulting from the short timescale under which rings are generated, in the order of picoseconds, for gigagauss magnetic-field strengths. Such short timescales make ring momentum structures highly resilient to diffusive processes such as magnetic-field curvature, guiding field inhomogeneities, and collisional effects. Conversely, for the case of laboratory conditions, the ring formation timescales and evolution are in the nanosecond timescale, and curvature or inhomogeneities in the magnetic field and the necessary plasma temperatures of $p_{\mathrm{th}}>m_e c$ are a challenge with state-of-the-art technology. Nonetheless, we conjecture that other radiatively cooled plasmas will also develop a population inversion, namely in laboratory conditions, e.g., high-energy particle beams undergoing betatron cooling are an ideal candidate to study analogous processes\cite{wang2002x,lu2007generating,davoine2018ion}. This will be presented in a future publication \cite{bilbaoinprpep2024}.

In conclusion, the full impact of radiation reaction cooling in the collective plasma dynamics has begun to be comprehended. The current results have applications for astrophysical processes, especially coherent maser radiation and firehose magnetic-field amplification. We conjecture that these are the first examples and that further work on different electromagnetic-field configurations will find new collective plasma physics triggered under extreme plasma physics conditions in laboratory or astrophysical settings.

\begin{acknowledgments}
We would like to thank T. Adkins, R. Bingham, S. Bulanov, T. Grismayer, P. Ivanov, M. Lyutikov, A. Schekochihin, R. Torres, and V. Zhdankin for fruitful discussions. This work was partially supported by FCT (Portugal)—Foundation for Science and Technology under the Project X-maser (No. 2022.02230.PTDC) and Grant No. UI/BD/151559/2021; European Union‘s Horizon 2020 research and innovation programme under grant agreement No. 653782; the European Research Council (ERC)—2015-AdG Grant No. 695088—InPairs; and UK EPSRC project No. 2397188.
\end{acknowledgments}
\appendix
\section{Vlasov QED synchrotron operator\label{ap:qed}}
TTo study the effects of quantum synchrotron cooling on a distribution of particles, we can construct a radiative operator. The master equation models the evolution of a distribution function due to stochastic synchrotron emission \cite{neitz2013stochasticity, vranic2016quantum_m, niel2018quantum}
\begin{align}
    \left.\frac{\partial f}{\partial t} \right|_{RR} =& \int d\mathbf{p}_\gamma \mathcal{W}\left( \mathbf{p} +\mathbf{p}_\gamma, \mathbf{p}_\gamma \right) f(\mathbf{p}+\mathbf{p}_\gamma,t)\nonumber\\
    &- f(\mathbf{p}) \int d\mathbf{p}_\gamma \mathcal{W}\left( \mathbf{p}, \mathbf{p}_\gamma \right)\label{eq:ap_master},
\end{align}
where $\mathcal{W}\left( \mathbf{p}, \mathbf{p}_\gamma \right)$ is the emission rate of a photon with momentum $\mathbf{p}_\gamma$ by an electron with momentum $\mathbf{p}$. The emission rate, assuming that photon emission occurs parallel to the particle momentum, is given by \cite{ritus1985quantum, vranic2016quantum_m}
\begin{align}
    \mathcal{W}\left( \mathbf{p}, \mathbf{p}_\gamma \right) =& \frac{1}{\sqrt{3}\pi} \frac{e \alpha}{B_{Sc} c^2}\frac{1}{\gamma^2}\delta\left(\mathbf{\hat{p}}_\gamma-\mathbf{\hat{p}}\right)\nonumber\\
    \times& \left[ \int_\nu^\infty K_{5/3}(x) dx + \frac{\left(\mathcal{E}_\gamma/\gamma \right)^2}{1-\mathcal{E}_\gamma/\gamma}K_{2/3}(\nu)\right],\label{eq:ap_rate_QED}
\end{align}
where $K_n(x)$ is the modified $n$-th order Bessel function of second-kind (or Basset function), $\mathcal{E}_\gamma$ is the energy of the emitted photon in $m_e c^2$, $\gamma$ is the Lorentz factor of the electron, and $\nu = 2(\mathcal{E}_\gamma/\gamma)/\left[3 \chi (1-\mathcal{E}_\gamma/\gamma)\right]$. Recall that $\chi$ is the Lorentz invariant quantity $\chi=p_\perp \left|\mathbf{B}\right| /(m_e B_\mathrm{Sc})$, that for a constant magnetic field simplifies to $\chi=p_\perp B/(m_e c B_{Sc})$. One notes that in the classical limit when the photon energy is much smaller than the kinetic energy of the electron, \emph{i.e.}, $(\mathcal{E}_\gamma/\gamma)^2 \ll1$, one recovers the classical radiation rate from the classical synchrotron, which can be derived from classical electrodynamics \cite{schwinger1949classical,schwinger2019classical,befki1966radiation, lightman1982relativistic}
\begin{equation}
    \mathcal{W}\left( \mathbf{p}, \mathbf{p}_\gamma \right) = \frac{\sqrt{3}}{2\pi} \frac{e^2 \alpha}{B_{Sc} c^2}\frac{\delta\left(\mathbf{\hat{p}}_\gamma-\mathbf{\hat{p}}\right) \omega_{ce}}{\omega_c}\int_{\omega/\omega_c}^\infty K_{5/3}(x) dx ,
\end{equation}
as $\nu=2\omega/(3\omega_{ce}\gamma^2)$, \emph{i.e.}, $\nu = \omega/\omega_c$ the ratio between the emitted frequency and critical frequency. This hints that the term proportional to $K_{2/3}(\nu)$ is of quantum nature.

We can expand the collision operator from Eq. (\ref{eq:ap_master}) following the standard procedure, \emph{i.e.}, with a Kramers-Moyal expansion \cite{pawula1967approximation, touati2014reduced, niel2018quantum}. To obtain a transport and diffusion operator
\begin{equation}
    \left.\frac{\partial f}{\partial t} \right|_{RR} = \nabla_p \cdot \left[\mathbf{F} f\right] + \frac{1}{2}\nabla^2_p\left[D f\right],
\end{equation}
where 
\begin{align}
    \mathbf{F} =& \frac{2}{3}\frac{9\sqrt{3}}{8\pi}\frac{e\alpha\chi^2 \mathbf{\hat{p}}}{B_{Sc}c^2}\nonumber\\ &\times\int_0^\infty\left[\frac{2\nu^2 K_{5/2}(\nu)}{(2+3\nu\chi)^2}  + \frac{36\nu^3\chi^2K_{2/3}(\nu)}{(2+3\nu\chi)^4}\right],
\end{align}
and
\begin{align}
    D =& \frac{2}{3}\frac{e\alpha\chi^3 \gamma}{B_{Sc}c^2}\nonumber\\ &\times\int_0^\infty\left[\frac{2\nu^3 K_{5/2}(\nu)}{(2+3\nu\chi)^3} + \frac{54\nu^4\chi^5K_{2/3}(\nu)}{(2+3\nu\chi)^5}\right].
\end{align}
In the limit where $\chi\lesssim1$, $\mathbf{F}$ and $D$ simplify to
\begin{equation}
    \mathbf{F} = \frac{2}{3}\frac{9\sqrt{3}}{8\pi}\frac{e\alpha\chi^2 \mathbf{\hat{p}}}{B_{Sc}c^2},\label{eq:ap_simpF}
\end{equation}
and
\begin{equation}
    D = \frac{55}{24\sqrt{3}}\frac{e\alpha\chi^3 \gamma}{B_{Sc}c^2}.\label{eq:ap_simpD}
\end{equation}
Note that Eq. (\ref{eq:ap_simpF}) yields the negative classical radiation reaction force in the relativistic limit of Eq. (\ref{eq:LLBfield_perp}). This yields the following collision operator
\begin{equation}
    \left.\frac{\partial f}{\partial t} \right|_{RR} = -\nabla_p \cdot \left[\mathbf{F}_{RR} f\right] + \frac{55}{48\sqrt{3}}\frac{e\alpha}{B_{Sc}c^2}\nabla^2_p\left[\chi^3\gamma f\right],
\end{equation}
where $\mathbf{F}_{RR}$ is the classical radiation reaction force (Eq. (\ref{eq:LLBfield})) in the relativistic limit $\gamma\gg1$. One can write the Vlasov equation for a ``semi-classical" synchrotron cooled plasma, where a diffusion operator includes the QED effects
\begin{multline}
    \frac{\partial f}{\partial t}+ \frac{\mathbf{p}}{\gamma} \cdot \mathbf{\nabla}_r f+ \mathbf{\nabla}_p \cdot \left[ \left( \textbf{F}_{RR}+ \textbf{F}_{L} \right) f \right]= \\ \frac{55}{48\sqrt{3}}\frac{e\alpha}{B_{Sc}c^2}\nabla^2_p\left[\chi^3\gamma f\right]. \label{eq:Vlasov_QED}
\end{multline}
The only remaining quantum term is in the right-hand-side of Eq. (\ref{eq:Vlasov_QED}), this can be seen by expanding $\chi=(B/B_{Sc}) (p_\perp/m_e c)= (\hbar e p_\perp)(m_e^3 c^3)$ and as $\alpha/B_{Sc} = 1/B_c$, where $B_c$ is the critical field in classical electrodynamics. Thus, the right-hand-side of Eq. (\ref{eq:Vlasov_QED}) is $\propto\hbar^3$. Thus, as the system becomes classical $\hbar\to0$ the Vlasov equation recovers the same form as Eq. (\ref{eq:pre_Vlasov}).

We have formally shown that classical radiation reaction is recovered in the regime $\chi\lesssim1$, and justified the use of Eq. (\ref{eq:pre_Vlasov}). This aligns with preliminary quantum particle-in-cell simulation results (See supplementary material Ref. \cite{bilbao2022radiation}). Moreover, as the distribution cools down the diffusive term in the right-hand-side of Eq. (\ref{eq:Vlasov_QED}) will become negligible as $\chi \to 0$ during the cooling process. By deriving Eq. (\ref{eq:Vlasov_QED}) we can begin to study the effects of QED synchrotron cooling in the semi-classical regime. This will be expanded in future work to quantify the effects of the diffusive term.

\bibliography{main}
\end{document}